# ICEV dismantling or recycling on a challenging environment

Manuel Alberto M. Ferreira [1], José António Filipe [2], José Moleiro Martins [3 *], Mário Nuno Mata [4], Pedro Neves Mata [5]

1. Instituto Universitário de Lisboa (ISCTE-IUL), ISTAR-IUL, Business Research Unit (BRU-IUL), Lisboa, Portugal; manuel.ferreira@iscte-iul.pt;
2. Instituto Universitário de Lisboa (ISCTE-IUL), ISTAR-IUL, Business Research Unit (BRU-IUL), Lisboa, Portugal; jose.filipe@iscte-iul.pt;
3. Instituto Politécnico de Lisboa; Instituto Universitário de Lisboa (ISCTE-IUL), Business Research Unit (BRU-IUL), Lisboa, Portugal; zdmmartins@gmail.com; *Corresponding Author
4. Instituto Politécnico de Lisboa; mnmata@iscal.ipl.pt;
5. Instituto Politécnico de Lisboa; pedro_mata@iscte-iul.pt;

**Abstract:** Nowadays Sustainability is a huge issue. Sustainability deals with the need for the protection of the natural environment and ecosystems health and requires innovation and commitment with the future. This manuscript uses the M|G|∞ queue system, modelling Internal Combustion Engine Vehicles (ICEV), normally cars but not only, which turn idle when conventional energy becomes scarce, or a new *status quo* is required. In such a case, they are recycled, becoming either EV-Electric Vehicles or HEV-Hybrid Electric Vehicles or FCEV-Fuel Cell Electric Vehicles, or are dismantled (DV-Dismantled Vehicles). Our model shows that when the rhythm ICEV become EV, HEV, FCEV and DV is greater than the rate at which they get idle the system tends to balance. In a cost-benefit analysis perspective, there are minimum benefits above which, both dismantling and recycling, are interesting. Additionally, the most interesting is the one for which the minimum benefit is the least.

## 1. Introduction

Sustainability is nowadays recognized as imperative, involving a vast and complex number of factors and having a relevant economic, environmental and social impact. Many causes for the deterioration of the environment make the situation unsustainable for numerous resources and ambiences. The environmental quality, for example - determinant for healthy communities - implies the need for clean air, healthy and balanced living resources, sustainable natural resources and a nontoxic environment. For instance, many health problems relate directly to air and water quality, to which Internal Combustion Engine Vehicles (ICEV) contribute.

The recent public health crisis that emerged from the spread of the Covid-19 disease made the collective conscience to recognize suddenly that the environmental map of the world got radically different, after pollution almost disappear from many regions of the world where previously was very serious. In fact, the world almost paralyzed and the traffic was radically reduced. The effect may seem like the one resulting from the replacement of ICEV for non-polluter vehicles.

In the present days, the collective urban consciousness, that is a product of the pandemics effects, invokes the need of reducing increasingly the levels of pollution up to the present levels, what remarkably will allow a very different quality in air, water and environment in general. Car manufacturers are modifying the production lines to work new non-polluter and less-polluter vehicles, changing the previous *status quo*. However, there are millions of cars already produced and an innovative solution may increase the speed at which the environmental quality improves.

Nowadays, this search for a sustainable performance is considered a core idea in business and society[1]. As the search for a balance between sustainability and economic development becomes indispensable, the problem of materials involved in cars composition, the problem of fuels in cars, or the alternative sources of energy are issues that need a strong discussion to allow fast decisions on the improvement processes, for getting the best options on cars production and on the viable reconversions.

Changes in the climatic conditions prompted by increased in pollution level indicate the sensitivity of natural and human systems, showing the urgency for new mentalities and needs. As there are many over-exploited non-renewable resources and a new paradigm is coming, innovation is required to overcome challenges posed by a coming new energy order. In the current days, conventional sources of energy such as oil, gas and coal continue to be broadly used, and in a near future, if the consumption remains at the present levels, they will end. Maybe firstly oil, then gas and lastly coal. Meanwhile it is interesting to notice that the consumption reduced drastically after the Covid-19 crisis, what made oil price to decrease for very low levels that were not seen since several years ago. It is evident that changes will drastically happen in the world and in collective conscience, and after this public health crisis ends, its effects will be much visible in social, economic and financial dimensions.

The relationship between sustainability and business turned into one of the principal debates at the national and international levels in both industrialized and emerging countries. Many investments around the world allow getting new technologies for alternatives to the currently most used sources of energy. New prospects for business are becoming evident and began to be experienced all over the world associated to the new sources of energy. Considering the viability of recovering the existing vehicles, which are moved by fossil fuels (based on oil, gas), we construct different scenarios to analyze the best options to give a new life to these vehicles.

We promote the idea that resources must be recycled and reused, as much as possible. A circular economy is nowadays considered vital in the businesses' systems; and trying, as much as possible, to create close-looped systems, minimizing the use of new resources and the wastes, carbon emissions, pollution. In our model, we propose means for studying the best alternatives within the automobile sector for recycling or dismantling, offering anyway the opportunity of new uses for resources. Solutions proposed by the model allow to keep the equipment for a longer period of utilization, when possible; or to reuse the components in alternative applications.

In this study, we apply a model to choose the most efficient technique, by starting from the technology and the process selected in order to make the conversion from a situation to another one. The goal of this paper is to show that Internal Combustion Engine Vehicles (ICEV), based on oil, in a collapsing scenario of the conventional energy or by considering a different energy paradigm can be converted; or can be simply dismantled. In this model, infinite servers' queues are used allowing to state that too many ICEV will become idle if either conventional energy misses or conventional energy becomes replaced by a renewable one. ICEV dismantling or recycling will become very usual once there will not be a way to get them functional with conventional oil, from the instant it is depleted or their use, in the existing conditions, is not allowed anymore.

The cadence at which we perform the recycling and dismantling actions is relevant, being central in this analysis the hazard rate function of the service time:

$$h(t) = \frac{g(t)}{1 - G(t)} \quad (1.1)$$

In (1.1), $G(t)$ is the distribution function and $g(t)$ is the probability density function[2].

The hazard rate function of the service time depends basically on the technology and protocols used in ICEV recycling and dismantling. Accordingly, the choice among the available ones is now of importance. The problem associated to the future of this kind of cars, in this case, comes in a very short time. The major producers reported that they are considering a decision on this in a very short time. They will stop making diesel cars very soon, for ethical and environmental reasons, still considering manufacturing gasoline cars in a middle term, and electric, hybrid and fuel cell cars.

Considering that the car industry assumes this awareness, ending the manufacture of diesel cars worldwide and implementing the manufacture of electric and hybrid cars, so a new era will come soon. Depending on how long rests the conventional energy, the question about the urgency of dismantling or recycling ICEV is posed. Also, a preliminary replacement of diesel cars by either electric or hybrid or fuel cell cars is considered. On this, from the economic point of view, it is difficult to predict what will happen because there are several variables to consider, for example: This decreed obsolescence of diesel cars makes them loosing value. A question may be also posed: alternatively, will there be a moment on which their value increases due to their increasing rarity? How costly will be manufacturing plants for dismantling or converting the existing diesel cars? What will occur with the diesel cars in the meantime by means of getting idle: dismantling, recycling? And, at what cost? Reducing the number of diesel cars possibly leads to a decrease in oil consumption, delaying the arrival of the traditional energy shortage situation.

This study aims mainly to contribute to the sustainability of the humankind standard of living, and a suitably preserved natural environment. In some reflections were when considering the need of reconversion in a situation of scarce oil environment.

Queues theory studies mathematically the queues, or the waiting lines. A model is created to predict queue length and the waiting time. Generally considered as a branch of operations research, queues theory contributes for the decision-making in business and a set of other very different areas. Some key moments will be presented here on Queues theory history. Queuing theory began in the first decade of the 20th Century, having been a work of the first publication on this area, applying a model to telecommunications[3]. In this paper, Erlang modeled the number of telephone calls arriving at an exchange by a Poisson process; in 1917 he would solve the M/D/1 queue; and later, in 1920, solved the M/D/k queueing model. Pollaczek solved the M/G/1 queue. A solution in probabilistic terms is now known as the Pollaczek–Khinchine formula[4]. Kleinrock, having first applied the queue theory to the message switching (1960's), worked later the queueing theory application to the packet switching (1970s) – his work is produced for example in Kleinrock[5].

The matrix geometric method and matrix analytic methods have allowed queues with phase-type distributed inter-arrival and service time distributions to be considered[6]. Problems such as performance metrics for the M/G/k queue remain an open problem[7]. These are some of the important works that have marked the evolution of the Queueing theory. Although many works have been developed along more than a century, the presented above are usually considered within the group of the ones that have been considered fundamental for the development of the theory.

The queue model here used is the M|G|∞ queue. The symbol ∞ means that it is a queue with infinite servers. This quality makes the model suitable to deal with great populations, as it is the case of the car's population considered in this study. Note that when talking about infinite servers, it is not considered the physical presence of an infinite number of servers. The physical realization of this infinity can translate into

ensuring that whenever a client arrives at the system it always finds an available server. This is what happens in the case under study. Whenever an owner decides to transform his/her vehicle, he/she immediately finds someone available to provide the service. In case of having to wait, the waiting time is incorporated into the service time.

As we will see later, using this model we will obtain stability conditions of the system, in terms of the hazard rate function of the service time and the rate of arrivals to the system. Note that this function is widely used in reliability theory, and in this study it will depend on the technological characteristics of the process used in the transformation of vehicles. From the many others that have been developed for this area, we select a set of the M|G|∞ queue applications, some of which have been developed in areas as the unemployment, diseases, finance, management, and logistics.

## 3. Materials and Methods

*Outlining the Model*

In the M|G|∞ queue, customers arrive according to a Poisson process at rate $\lambda$ and each one receives a service which time is a positive random variable with distribution function $G(\cdot)$ and mean value $\alpha$. There are infinite servers, that is: when a customer arrives, it always finds an available server. The service of a customer is independent from the other customers' services and from the arrivals process. An important parameter is the traffic intensity, called $\rho=\lambda\alpha$. The M|G|∞ queue has neither losses nor waiting.

In what concerns to the present study, the costumers are the ICEV that become idle. The arrivals rate is the rate at which the ICEV become idle. The service time for each one is the time that goes from the instant they get idle until the instant they are either recycled or dismantled.

The hazard rate function of the service time is the rate at which the services end. For the situation under study in this paper, is the rate at which the motorcars are either recycled, turning either EV or HEV, or dismantled, turning DV.

Denoting $p_{1'0}(t) = G(t)e^{-\lambda \int_0^t [1-G(v)]dv}$, the probability the M|G|∞ queue has no costumers at instant t, being the time origin an instant at which a costumer arrives at the system finding it empty (symbolized by the $1'$).

Proposition 3.1

If $G(t) < 1, t > 0$ continuous and differentiable and

$$h(t) \geq \lambda, t > 0 \quad (3.1)$$

$p_{1'0}(t)$ is non-decreasing.

Dem.: It is enough to note that $\frac{d}{dt} p_{1'0}(t) = e^{-\lambda \int_0^t [1-G(v)]dv}(1 - G(t))(h(t) - \lambda G(t))$.

Obs.:

-If the rate at which the services end is greater or equal than the costumers 'arrivals rate $p_{1'0}(t)$ is non-decreasing.

-For the M|M|∞ system, exponential service times, $h(t) = 1/\alpha$ and (3.1) is equivalent to

$$\rho \leq 1 \quad (3.2).$$

Either Equation (3.1) evidences that if the recycling or the dismantling rate is greater or equal than the rate at which the motorcars become idle, the probability that the system is empty at instant t, meaning it that there is no idle ICEV, does not decrease with t. Therefore, the system has a tendency to balance as far as time goes on.

Denoting now $\mu(1',t) = 1 - G(t) + \lambda \int_0^t [1 - G(v)] dv$ the mean number of customers in the M|G|∞ queue at instant t, being the time origin an instant at which a costumer arrives at the system finding it empty (symbolized by the $1'$).

Proposition 3.2

If $G(t) < 1, t > 0$ continuous and differentiable and

$$h(t) \geq \lambda, t > 0 \quad (3.3)$$

$\mu(1',t)$ is non-increasing.

Dem.: It is enough to note that $\frac{d}{dt}\mu(1',t) = (1 - G(t))(\lambda - h(t))$.

Obs.:

-If the rate at which the services end is greater or equal than the customer's arrivals rate, $\mu(1',t)$ is non-increasing.

-For the M|M|∞ system $h(t) = 1/\alpha$ and (2.3) is equivalent to

$$\rho \leq 1 \quad (3.4).$$

Either equation (3.3) evidences that if the recycling or the dismantling rate is greater or equal than the rate at which the ICEV become idle, the mean number of ICEV in the system does not increase with time. This means that the system has a propensity to balance as far as time goes on.

## 4. Results and Discussion

*Performing cost-benefit analysis*

In the former section, we saw how important the roles of h (t) were and $\lambda$, in monitoring the ICEV recycling and dismantling management.

To perform an economic analysis, based on the model presented behind, consider additionally p as the probability, or percentage, of the ICEV arrivals designed to the recycling being consequently 1-p the same to the dismantling. In addition, be q the percentage or probability of ICEV designed for recycling that turn EV; r will be the same for ICEV designed for recycling that turn HEV; and 1-q-r will be the same for ICEV designed for recycling that turn FCEV. Call $h_i(t), c_i(t)$ and $b_i(t), i = EV, HEV, FCEV, DV$ the hazard rate function, the mean cost and the mean benefit, respectively for an ICEV turn either EV or HEV or FCEV or DV. Therefore, the total cost per unit of time for motor cars recycling and dismantling is:

$$C(t) = \lambda[pqc_{EV}(t) + prc_{HEV}(t) + p(1 - q - r)c_{FCEV}(t) + (1 - p)c_{DV}(t)] \quad (4.1)$$

and the benefit per unit of time resulting from recycling and dismantling

$$B(t) = b_{EV}(t)h_{EV}(t) + b_{HEV}(t)h_{HEV}(t) + b_{FCEV}(t)h_{FCEV}(t) + b_{DV}(t)h_{DV}(t) \quad (4.2).$$

So consider a period of time T. It must be $\int_0^T B(t)dt > \int_0^T C(t)dt$. It is not a simple matter to deal analytically with this expression. But, considering $b_{EV}(t), b_{HEV}(t),$

$b_{FCEV}(t)$ and $b_{DV}(t)$ are all constant in $[0,T]$ with values $b_{EV}, b_{HEV}, b_{FCEV}$ and $b_{DV}$, respectively, it is obtained:

- Recycling, turning ICEV in EV, is interesting, if

$$b_{EV} > \max\left\{\frac{\lambda[pqC_{EV}^T + prC_{HEV}^T + p(1-q-r)C_{FCEV}^T + (1-p)C_{DV}^T]}{\ln\frac{1-G_{EV}(0)}{1-G_{EV}(T)}} - \frac{b_{HEV}\ln\frac{1-G_{HEV}(0)}{1-G_{HEV}(T)} + b_{FCEV}\ln\frac{1-G_{FCEV}(0)}{1-G_{FCEV}(T)} + b_{DV}\ln\frac{1-G_{DV}(0)}{1-G_{DV}(T)}}{\ln\frac{1-G_{EV}(0)}{1-G_{EV}(T)}}, 0\right\} \quad (4.3)$$

- Recycling, turning ICEV in HEV, is interesting if

$$b_{HEV} > \max\left\{\frac{\lambda[pqC_{EV}^T + prC_{HEV}^T + p(1-q-r)C_{FCEV}^T + (1-p)C_{DV}^T]}{\ln\frac{1-G_{HEV}(0)}{1-G_{HEV}(T)}} - \frac{b_{EV}\ln\frac{1-G_{EV}(0)}{1-G_{EV}(T)} + b_{FCEV}\ln\frac{1-G_{FCEV}(0)}{1-G_{FCEV}(T)} + b_{DV}\ln\frac{1-G_{DV}(0)}{1-G_{DV}(T)}}{\ln\frac{1-G_{HEV}(0)}{1-G_{HEV}(T)}}, 0\right\} \quad (4.4)$$

- Recycling, turning ICEV in FCEV, is interesting, if

$$b_{FCEV} > \max\left\{\frac{\lambda[pqC_{EV}^T + prC_{HEV}^T + p(1-q-r)C_{FCEV}^T + (1-p)C_{DV}^T]}{\ln\frac{1-G_{FCEV}(0)}{1-G_{FCEV}(T)}} - \frac{b_{EV}\ln\frac{1-G_{EV}(0)}{1-G_{EV}(T)} + b_{HEV}\ln\frac{1-G_{HEV}(0)}{1-G_{HEV}(T)} + b_{DV}\ln\frac{1-G_{DV}(0)}{1-G_{DV}(T)}}{\ln\frac{1-G_{FCEV}(0)}{1-G_{FCEV}(T)}}, 0\right\} \quad (4.5)$$

- Dismantling is interesting if

$$b_{DV} > \max\left\{\frac{\lambda[pqC_{EV}^T + prC_{HEV}^T + p(1-q-r)C_{FCEV}^T + (1-p)C_{DV}^T]}{\ln\frac{1-G_{DV}(0)}{1-G_{DV}(T)}} - \frac{b_{EV}\ln\frac{1-G_{EV}(0)}{1-G_{EV}(T)} + b_{HEV}\ln\frac{1-G_{HEV}(0)}{1-G_{HEV}(T)} + b_{FCEV}\ln\frac{1-G_{FCEV}(0)}{1-G_{FCEV}(T)}}{\ln\frac{1-G_{DV}(0)}{1-G_{DV}(T)}}, 0\right\} \quad (4.6)$$

where $C_i^T = \int_0^T c_i(t)dt, i = EV, HEV, FCEV, DV$.

But if moreover $G_{EV}(t), G_{HEV}(t), G_{FCEV}(t)$ and $G_{DV}(t)$ are all exponential, with means $\alpha_{EV}, \alpha_{HEV}, \alpha_{FCEV}$ and $\alpha_{DV}$, respectively, (4.3), (4.4), (4.5) and (4.6) become:

- Recycling, turning ICEV in EV, is interesting, if

$$b_{EV} > \max\left\{\frac{\rho_{EV}[pqC_{EV}^T + prC_{HEV}^T + p(1-q-r)C_{FCEV}^T + (1-p)C_{DV}^T]}{T} - b_{HEV}\frac{\alpha_{EV}}{\alpha_{HEV}} - b_{FCEV}\frac{\alpha_{EV}}{\alpha_{FCEV}} - b_{DV}\frac{\alpha_{EV}}{\alpha_{DV}}, 0\right\} \quad (4.7)$$

with $\rho_{EV} = \lambda\alpha_{EV}$

- Recycling, turning ICEV in HEV, is interesting if

$$b_{HEV} > max\left\{\frac{\rho_{HEV}\left[pqC_{EV}^T + prC_{HEV}^T + p(1-q-r)C_{FCEV}^T + (1-p)C_{DV}^T\right]}{T}\right.$$

$$\left. - b_{EV}\frac{\alpha_{HEV}}{\alpha_{EV}} - b_{FCEV}\frac{\alpha_{HEV}}{\alpha_{FCEV}} - b_{DV}\frac{\alpha_{HEV}}{\alpha_{DV}}, 0\right\} \quad (4.8)$$

with $\rho_{HEV} = \lambda\alpha_{HEV}$

- Recycling, turning ICEV in FCEV, is interesting, if

$$b_{FCEV} > max\left\{\frac{\rho_{FCEV}\left[pqC_{EV}^T + prC_{HEV}^T + p(1-q-r)C_{FCEV}^T + (1-p)C_{DV}^T\right]}{T}\right.$$

$$\left. - b_{EV}\frac{\alpha_{FCEV}}{\alpha_{EV}} - b_{HEV}\frac{\alpha_{FCEV}}{\alpha_{HEV}} - b_{DV}\frac{\alpha_{FCEV}}{\alpha_{DV}}, 0\right\} \quad (4.9)$$

with $\rho_{FCEV} = \lambda\alpha_{FCEV}$

- Dismantling is interesting if

$$b_{DV} > max\left\{\frac{\rho_{DV}\left[pqC_{EV}^T + p(1-q)C_{HEV}^T + p(1-q-r)C_{FCEV}^T + (1-p)C_{DV}^T\right]}{T}\right.$$

$$\left. - b_{EV}\frac{\alpha_{DV}}{\alpha_{EV}} - b_{HEV}\frac{\alpha_{DV}}{\alpha_{HEV}} - b_{FCEV}\frac{\alpha_{DV}}{\alpha_{FCEV}}, 0\right\} \quad (4.10)$$

with $\rho_{DV} = \lambda\alpha_{DV}$.

Let´s illustrate formulae (4.7), (4.8), (4.9) and (4.10) application in a practical situation. Consider, for instance, a period of 30 days ( $T = 30$ ) and $\lambda = 30/day$. Also, $p = 0,6, q = 0,2$ and $r = 0,1$. In the following table we give a set of possible values for the parameters defined above:

Table 1. Parameters values in the example

| i | $\alpha_i$ | $b_i$ | $C_i^{30}$ |
|---|---|---|---|
| EV | 1 day | 50 euro | 150 euro |
| HEV | 2 days | 30 euro | 200 euro |
| FCEV | 2 days | 30 euro | 175 euro |
| DV | *1,5 days* | *20 euro* | *100 euro* |

Entering with these values in formulae (4.7), (4.8), (4.9) and (4.10) we obtain respectively:

- $50\ euro > 48{,}75\ euro$,
- $30$ euro $> 27{,}50$ *euro*,
- $30$ euro $> 27{,}50$ *euro*,
- $20$ euro $> 43{,}13$ (false inequality).

So, in this context, dismantling is the only option that is not economically interesting. Otherwise, the most efficient options are to recycle, transforming ICEV into HEV or FCEV, because they are the most economically interesting, from lower levels. Thus, this model can function as a decision support system.

This model can also be applicable to the dismantling and recycling situation resulting from the universal abandonment of the construction of diesel cars described in section 1. Now the option of recycling should consider a fourth option, the conversion of diesel cars to gasoline cars.

In our world, resources get scarcer each day and the collective *status quo* define new principles for life. The need of finding the best option for resources is posed. Considering the importance of developing strategies to preserve resources and to improve efficiency, this methodology allows the economic and financial monitoring of the studied situations, the technological choice of the process for each option and allows to note that the paces of recycling and dismantling must be adapted to the pace of energy extinguishing that is intended to be replaced (or simply the arrival of a new life collective option). Accordingly, this model allows to promote good practices and get efficient decisions, by defining innovative proposals in the context of the discussed problem in this paper.

As equipment has the possibility of recycling, reducing and minimizing waste is key for new economy and social life, within the circular economy, providing the best practices for the management of this type of resources. Vehicles can be studied in order to provide the best option within the decision-making processes, aiming sustainable decisions according to the principle of resources conservation and environmental protection.

The solutions proposed contribute for the reduction of resources wastes, and carbon emissions and pollution's minimization. Our study recommends alternative solutions based on the model, allowing consistent, valuable and considerable achievements in the field of sustainability and resources preservation. By suggesting the best choices for ICEV, when studying the scenarios of recycling or dismantling, the model offers the best applicable solution for the equipment.

In this situation, although this is not a problem, in general, given the high demand for these services, it is relevant to check if customers' arrivals are made according to a Poisson process. It is anyway important to estimate $\lambda$ and h(t) for results concerning the system given the available data. A right estimate of $\lambda$ depends on the arrivals process to be Poisson. The approach is to decide for a mean $\lambda$ estimate for a given period since it is easy to admit that the arrivals rate will depend on time. For very large populations, such as the studied ones, the estimation of h(t) is in general technically not easy. Considering that, it is convenient to estimate directly h(t) instead of estimating initially the service time distribution and then the consequent computation of h(t). All this is particularly easy for exponential service times once in this case, $h(t)$ does not depend on time.

## 6. Conclusions

This is a study that uses a model that shows the tendency to the balance of a system when the rhythm ICEV becomes EV, HEV, FCEV, and DV is bigger than the rate at which they become idle. In terms of a cost-benefit analysis, there are minimum benefits above which, from both dismantling and recycling, are interesting. And the most interesting is the one for which this minimum benefit is the least.

Our study gives indications for the development of efficient performances and for the support of the decision-making process. Innovative processes consubstantiate a new force for the implementation of solutions to problems that are faced before new paradigms and environmental contexts. Circular economy is nowadays catching much attention inside the automotive industry and it requires to be discussed. Our paper deals with this need and proposes ways to deal with. The discussion around it envisages to direct attention to the possibility of reducing the use of natural resources and eliminating wastes, providing the possibility of carrying out good practices in the resources' utilization.

The results show we can realize the importance of the proposed politics. As the automobile industry is one of the most active and dynamic industries of the world, and knowing that this sector manages a substantial set of resources areas and is the receptor of numerous subsectors' components, the solutions obtained allow to get efficient ways of dealing with the problem that the study approaches.

The model developed in this paper contributes for an improved analysis of this kind of problems. With possible modifications, for some other social problems (as unemployment, health, pensions' funds, investment projects or repair systems) this model is also appropriate.

**Acknowledgments:** The authors thank the Instituto Universitário de Lisboa and ISTAR-IUL, for their support, under the project UID/Multi/04466/2019.

This work was partially supported by the Polytechnic Institute of Lisbon through the Projects for Research, Development, Innovation and Artistic Creation (IDI&CA), within the framework of the project IEOMAB—internationalization of companies operating in the Angolan and Brazilian markets, IPL/2019/IEOMAB_ISCAL.